\documentclass[]{pasj01}
\Received{2019/12/19}%{yyyy/mm/dd}
\Accepted{2020/03/17}%{yyyy/mm/dd}
\begin{document} 

\title{FOREST unbiased Galactic plane imaging survey with the Nobeyama 45 m telescope (FUGIN). VII. molecular fraction of HI clouds}
\author{Hiroyuki \textsc{Nakanishi}\altaffilmark{1}}%
\altaffiltext{1}{Graduate School of Science and Engineering, Kagoshima University, 1-21-35 Korimoto Kagoshima}
\email{hnakanis@sci.kagoshima-u.ac.jp}
\author{Shinji \textsc{Fujita},\altaffilmark{2}}
\altaffiltext{2}{Department of Astrophysics, Nagoya University, Furo-cho, Chikusa-ku, Nagoya, Aichi 464-8602, Japan}
\author{Kengo \textsc{Tachihara},\altaffilmark{2}}
\author{Natsuko \textsc{Izumi}\altaffilmark{3}}
\altaffiltext{3}{College of Science, Ibaraki University, Bunkyo 2-1-1, Mito, Ibaraki, 310-8512, Japan}
\author{Mitsuhiro \textsc{Matsuo}\altaffilmark{4}}
\altaffiltext{4}{Nobeyama Radio Observatory, National Astronomical Observatory of Japan, National Institutes of Natural Sciences, 462-2 Nobeyama Minamimaki, Minamisaku, Nagano 384-1305, Japan}
\author{Tomofumi \textsc{Umemoto}\altaffilmark{4}}
\author{Yumiko \textsc{Oasa}\altaffilmark{5}}
\altaffiltext{5}{Faculty of Education, Saitama University, 255 Shimo-Okubo, Sakura, Saitama, Saitama 338-8570, Japan}
\author{Tsuyoshi \textsc{Inoue}\altaffilmark{2}}

%%% end:list of authors

%% `\KeyWords{}' always has to be placed before ``\maketitle'' 
%%  List of Key Words:  https://academic.oup.com/pasj/pages/Pasj_Keywords 

\maketitle

\begin{abstract}
In this study, we analyze molecular gas formation in neutral atomic hydrogen (HI) clouds using the latest CO data obtained from the four-beam receiver system on a 45-m telescope (FOREST) unbiased Galactic plane imaging survey with the Nobeyama 45-m telescope (FUGIN) and HI data taken from the Very Large Array (VLA) Galactic plane survey (VGPS). We applied a dendrogram algorithm to the HI data cube to identify HI clouds, and we calculated the HI mass and molecular gas mass by summing the CO line intensity within each HI cloud. On the basis of the results, we created a catalog of 5,737 identified HI clouds with local standard of rest (LSR) velocity of $V_{\rm LSR}\le -20$ km s$^{-1}$ in Galactic longitude and latitude ranges of $20^\circ \le l \le 50^\circ$ and $-1^\circ \le b \le 1^\circ$, respectively. 
We found that most of the HI clouds are distributed within a Galactocentric distance of 16 kpc, most of which are in the Cold Neutral Medium (CNM) phase. In addition, we determined that the high-mass end of the mass HI function is well fitted with the power-law function with an index of 2.3. Although two sequences of self-gravitating and diffuse clouds are expected to appear in the M$_{\rm tot}$-M$_{{\rm H}_2}$ diagram according to previous works based on a plane-parallel model, the observational data show only a single sequence with large scattering within these two sequences. This implies that most of the clouds are mixtures of these two types of clouds. Moreover, we suggest the following scenario of molecular gas formation: An HI-dominant cloud evolved with increasing H$_2$ mass along a path of $M_{{\rm H}_2} \propto M_{\rm tot}^2$ by collecting diffuse gas before reaching and moving along the curves of the two sequences. 
\end{abstract}

\section{Introduction}
Neutral atomic hydrogen (H1) and molecular hydrogen (H$_2$) constitute the main components of the Galactic interstellar medium (ISM). The former can be observed directly in the 1420 MHz radio line, whereas the latter is generally traced by radio lines of another molecular species such as CO \citep{nak03, nak06a, nak16}. Although the (H$_2$) gas traced by the CO line is observationally found in dense gas regions, the HI line is a tracer of the diffuse gas region.

The fraction of molecular component to the total gas is often defined as $f_{\rm mol} = \Sigma_{{\rm H}_2}/(\Sigma_{\rm HI}+\Sigma_{{\rm H}_2})$, where $\Sigma_{\rm HI}$ and $\Sigma_{{\rm H}_2}$ are the HI and H$_2$ surface densities, respectively. It is observationally shown that the molecular fraction $f_{\rm mol}$ is high near the Galactic center and decreases with the Galactocentric distance \citep{sof95,sof16}.  The molecular fraction is considered to be a useful tool for examining the physical condition of the interstellar medium because it depends on the pressure ($P$), ultraviolet (UV) radiation field ($U$), and the metallicity ($Z$), as theoretically suggested \citep{elm93,kru08,mck10,ste14}.  

According to \citet{elm93}, the molecular fraction in the Galactic scale can be calculated by assuming that interstellar clouds are classified into diffuse and self-gravitating ones and by summing all of the molecular components in individual clouds, considering that the cloud mass function follows a power law. Because the gas density increases with the interstellar pressure $P$ and the shielding effect increases with the metallicity $Z$, the molecular fraction increases with $P$ and $Z$. However, the molecular fraction decreases with $U$ because the molecular gas is photo-dissociated by UV photons. It should be noted that the molecular fraction is sensitive to the metallicity $Z$ because the metallicity is proportional to the amount of ISM dust, which also works as a catalyst for forming H$_2$ from HI gases \citep{hon95}. 

These theoretical models closely match the observational values at the kiloparsecs scale \citep{hon95,nak06b, kru09,tan14,sof16}. In these models, the molecular component forms a core at the center of each cloud that is surrounded by an HI layer, which shields the central molecular core from the photo-dissociation owing to the UV radiation field \citep{elm93,kru08,mck10,ste14}. Such models of ISM clouds are often referred to as plane-parallel photo-dissociation region (PDR) models \citep{tie85,hol97}, which also suggest existence of [CII] and [CI] layers surrounding an H$_2$ core in addition to HI gas. 

However, such a plane-parallel model appears to be overly simplified. Recent [CI] and [CII] observations \citep{kam03,kra08,shi13} suggest that this model cannot completely explain the [CI] and [CII] data, which instead imply that the CO gas is in a phase of clumpy cloudlets, as suggested by \citet{tac12}. Therefore, it is necessary to verify whether most of the ISM clouds can be explained with the plane-parallel model based on observational data. 

For this purpose, a wide spatial dynamic range with high resolution and a wide field of view is essential. We used CO and HI data recorded during the four-beam receiver system on a 45-m telescope (FOREST) unbiased Galactic plane imaging survey with the Nobeyama 45-m telescope (FUGIN) and Very Large Array (VLA) Galactic plane survey (VGPS) projects \citep{ume17,sti06}. These data have a large spatial dynamic range and are ideal for evaluating whether the atomic and molecular components are distributed inside an HI cloud, as shown in the simple plane-parallel model described in previous research. 

As part VI of the FUGIN project series, we report on a study of the molecular gas formation by comparing CO and HI data. Sections 2 and 3 summarize the data and data analysis, respectively, and section 4 details the process of obtaining the results. In section 5, we discuss the properties of identified HI clouds such as HI cloud distribution as a function of Galactocentric distance, mass function, and H$_2$ gas distribution and fraction in each clouds. Finally, section 6 provides a summary of the conclusions.

\section{Data}
The $^{12}$CO($J=1$--0) data tracing the H$_2$ were obtained through the FUGIN  project, which covered Galactic longitude ranges of $10^\circ \le l \le 50^\circ$ and $198^\circ \le l \le 236^\circ$ and a latitude range of $-1^\circ \le b \le 1^\circ$ in $^{12}$C$^{16}$O, $^{13}$C$^{16}$O, and $^{12}$C$^{18}$O $J = 1$--0 lines \citep{ume17}. 

In this study, we focus on the $^{12}$C$^{16}$O data in the longitude range of $10^\circ \le l \le 50^\circ$ to compare the molecular fraction with the high-resolution HI data. The angular resolution was 20′′, whereas the angular sampling was 8.5′′. The sensitivity in the antenna temperature $T_{\rm A}^*$ was 0.24 K for the velocity resolution of 1.3 km s$^{-1}$. The antenna temperature was converted into the main beam temperature using a main beam efficiency of 0.43.   

The HI data were taken from the archived data of the VGPS, in which a high-resolution HI survey was conducted with the VLA covering a Galactic longitude range of $18^\circ \le l \le 67^\circ$ and a latitude range of $|b| \le 1.3^\circ$ or $|b| \le 2.3^\circ$. The angular resolution was 1', and the velocity resolution was 1.56 km s$^{-1}$ \citep{sti06}. The sensitivity was 2 K per channel width of 0.824 km s$^{-1}$. The missing flux was recovered by single dish observations conducted using a Green Bank 100-m telescope. 

\begin{figure*}[h]
\begin{center}
\rotatebox{0}{\includegraphics[width=190mm]{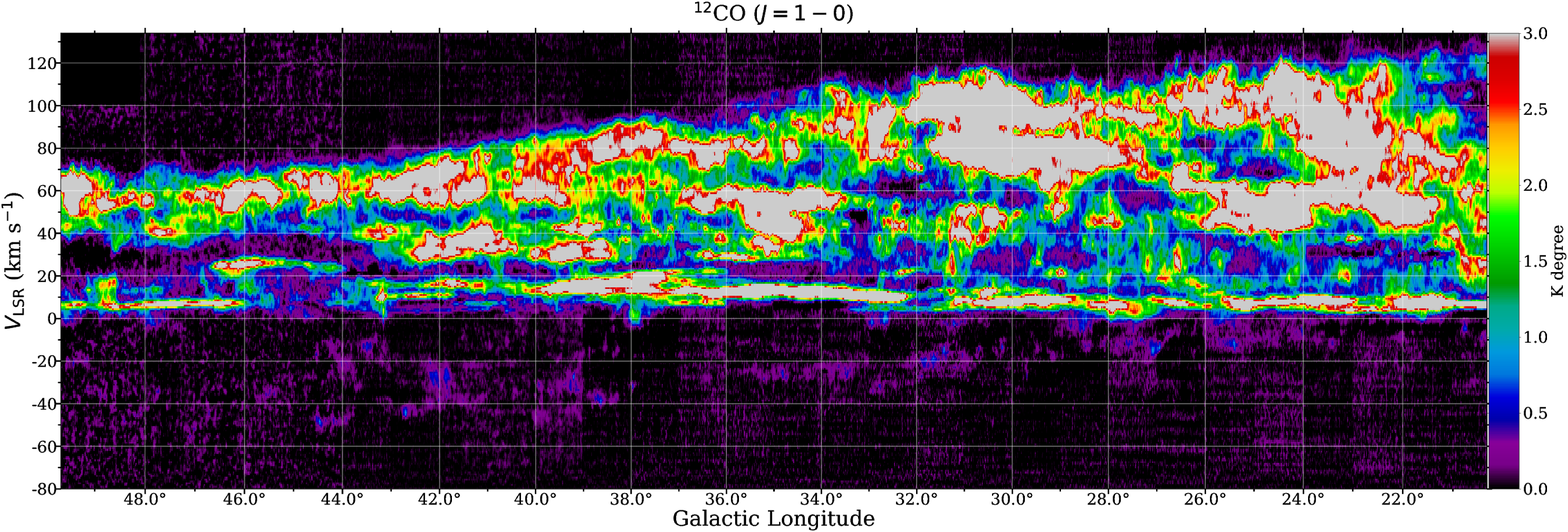}}
\rotatebox{0}{\includegraphics[width=190mm]{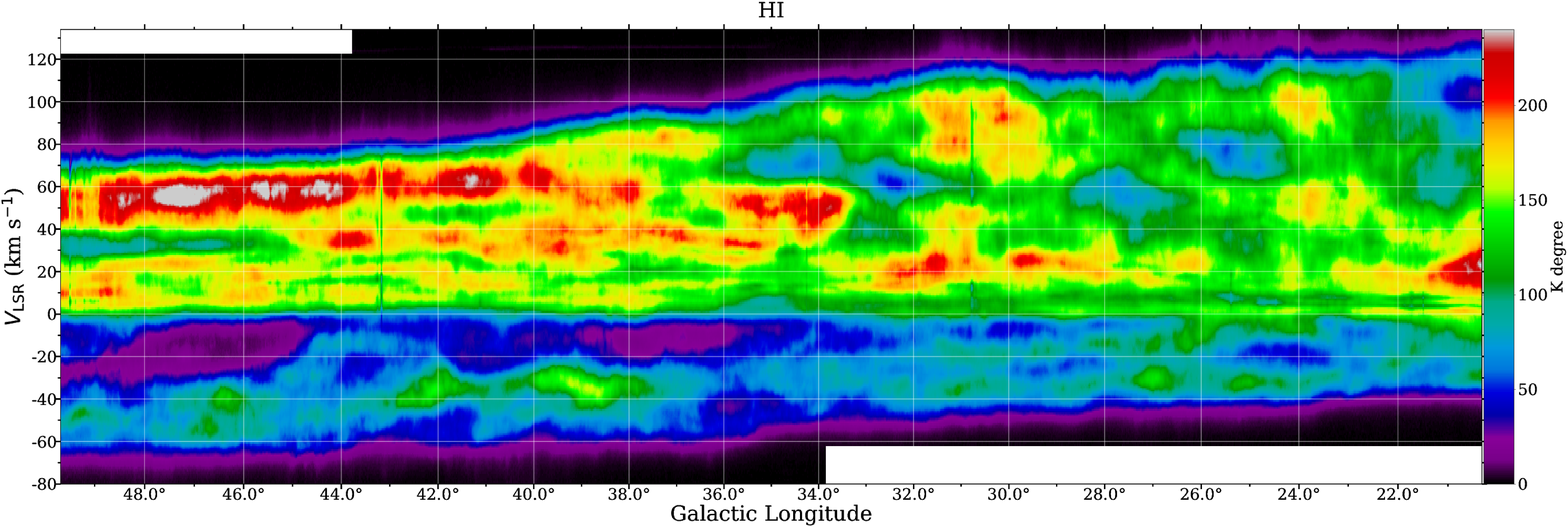}}
\end{center}
  \caption{Longitude-velocity diagrams of $^{12}$CO($J=1$--0) (top) and HI (bottom) data obtained from FUGIN and VGPS, respectively. The data were integrated in the Galactic latitude ranges of  $|b| \le 1.0^\circ$ for CO and $|b| \le 1.3^\circ$ for HI. This study focuses on the regions of $|b| \le 1.0^\circ$ and $V_{\rm LSR}\le -20$ km s$^{-1}$.} \label{fig:lv}
\end{figure*}

Both data were regridded so that the pixel size and velocity spacing were set at 18'' and 0.824 km s$^{-1}$ for comparison. {Because the CO data were convolved so that the convoluted beam size matched that of the HI data (1'), the sensitivity of the convolved CO data became $\sim 0.2$ K ($^{12}$C$^{16}$O) and $\sim 0.1$ K ($^{13}$C$^{16}$O and $^{12}$C$^{18}$O). }
The longitude-velocity diagrams of CO and HI data are shown in the top and bottom panels of figure \ref{fig:lv}, respectively.
 
\section{Data Analysis}
First, we applied the dendrogram algorithm developed by \citet{ros08} to the VGPS HI data cube in order to identify the HI clouds. To avoid near-far ambiguities in the kinematic distances of clouds inside the Solar circle, we analyzed clouds having only negative LSR velocities. Also considering that the velocity dispersion is about 10 km s$^{-1}$ \citep{mal95}, we restricted our analysis to $V_{\rm LSR} \le -20$ km s$^{-1}$ to be safe. Cloud identification was conducted for subsets of data divided into those with a longitude width of $10^\circ$. Neighboring subsets were overlapped by $5^\circ$, and double-counted HI clouds were merged into one. The dendrogram outputs bunches of voxels surrounded by the minimum contour level $T_{\rm min}$ as trunks and local maxima as leaves  (``voxel'' means three dimensional pixel in $l$, $b$, and $V_{\rm LSR}$ axes). The minimum contour level $T_{\rm min}$ was set to be 5 K, and the minimum voxel number was 125.   
We define ``leaf'' output by the dendrogram as ``HI cloud'' in this paper. It should be noted that this kind of segmentation method such as dendrogram algorithm cannot recover all the HI emission including diffuse components as is mentioned in the next section because spatially extended emissions with broad line widths are missed.

The heliocentric distances $D$ of the identified HI clouds were calculated on the basis of the kinematic distance by using the same rotation curve as that adopted by \citet{nak16}. Since the velocity of an HI cloud in the Local Standard of Rest (LSR) $V_{\rm LSR}$ is calculated as
\begin{equation}
V_{\rm LSR} = \left( {R_0\over R} V(R) - V_0\right)\sin{l}\cos{b}, 
\end{equation}
where $R$ and $V(R)$ denote the Galactocentric distance and the rotation curve, respectively, if the HI disk is assumed to circularly rotate. The subscript 0 means the values at the Sun. Given that the Galactic longitude is $l$, the Galactocentric distance $R$ can be calculated with the heliocentric distance $D$ as 
\begin{equation}
R=\sqrt{D^2+R_0^2-2DR_0 \cos{l}}.  
\end{equation}
Therefore, the heliocentric distance $D$ of an HI cloud can be calculated with the LSR velocity $V_{\rm LSR} (D)$, which can be written as a function of $D$ by eliminating $R$ from equations (1) and (2).

The inner and outer rotation curves were taken from \citet{cle85} and \citet{deh98}, respectively. The Galactic constants, $R_0$ (the Galactocentric distance of the Sun) and $V_0$ (the rotational velocity at the Sun), were 8 kpc and 217 km s$^{-1}$, respectively, as adopted by \citet{deh98} and \citet{nak16}. The masses of the clouds were calculated by multiplying the sum of the brightness temperature $\Sigma T_{\rm HI}$ with the pixel size $\theta_{\rm pixel}=18''$, velocity resolution $\Delta v=0.824$ km s$^{-1}$, the conversion factor from HI integrated intensity to HI column density $X_{\rm HI}=1.8 \times 10^{18}$ H cm$^{-2}$ (K km s$^{-1}$)$^{-1}$, and the square of the heliocentric distance $D$. 
In order to estimate the H$_2$ mass included in each HI cloud, we summed the brightness temperature of the $^{12}$CO data cube within each HI cloud. By including voxels with brightness temperatures lower than the noise level, diffuse emission was recovered in the same manner as that using stacking analysis \citep{mor15}. The error was estimated by multiplying the noise in the brightness temperature with the square root of number of voxels. For clouds in which the H$_2$ mass was less than three times the root–mean-square (RMS) noise, the  H$_2$ mass was not considered entered. The CO-to-H$_2$ conversion factor was set at $X_{\rm CO} =1.8\times 10^{20}$ H$_2$ cm$^{-2}$ K$^{-1}$ (km s$^{-1}$)$^{-1}$, as adopted from \citet{dam01}.

\section{Results}
The number of identified HI clouds was 5,737, some of which are listed in table 1, where Column (1) shows the cloud identification number; Columns (2) and (3) show the Galactic longitude and latitude, respectively; Column (4) shows the local standard of rest (LSR) velocity; Column (5) shows the heliocentric distance; Column (6) shows the Galactocentric distance; Column (7) shows the major and minor axis radii; Column (8) shows the position angle; Column (9) shows the velocity dispersion; Column (10) shows the HI mass; and Column (11) shoes the H$_2$ mass. A complete list of these parameters is available online as digital data.  

The maximum and minimum values of the HI clouds mass were $1.35\times 10^4$ M$_\odot$ and $14.0$ M$_\odot$, respectively. The mean mass was calculated to be $5.59\times 10^2$ M$_\odot$. The nearest and farthest HI clouds were located at the heliocentric distances of 11.9 kpc and 25.1 kpc, respectively. The mean heliocentric distance was 16.4 kpc, which can be regarded as the typical distance of the HI clouds in our sample. The corresponding minimum, maximum, and mean values of the Galactocentric distances were 9.1 kpc, 18.2 kpc, and 11.3 kpc, respectively.

Considering that minimum voxel number of 125 and minimum contour level of $T_{\rm min}=5$ K were adopted as described in section 3 and that the maximum heliocentric distance was $D_{\rm max}=25.1$ kpc, HI clouds with mass less than $M=125 X_{\rm HI} T_{\rm min} \Delta v (D_{\rm max} {18\over 3600} {\pi\over 180})^2 {m_{\rm p}\over M_\odot}=36$ M$_\odot$ were not fully identified. 
%125x1.82e18*5*0.824*(25.1e3*3.09e18*18/3600*3.14159/180)^2*1.67e-27/1.989e30=1.25*1.82*5.0*0.824*1.67/1.989*10^(2+18-27-30)(2.51*3.09*18/36*3.14159/18)^2*10^((4+18-2-1)*2)=7.8697e-37 * 0.458e38=36Mo

Similarly, because the minimum heliocentric distance of the HI clouds was 11.9 kpc, the mass of the faintest detectable cloud was estimated to be as small as 8.1 M$_\odot$

The maximum, minimum, and mean values of the mass of H$_2$ gas were $1.37\times 10^4$ M$_\odot$, $0.300$ M$_\odot$, and $2.84 \times 10^2$ M$_\odot$, respectively. The H$_2$ gas was detected among HI clouds in the heliocentric distance range of 11.9--25.1 kpc. The mean heliocentric distance was 16.4 kpc, which is almost the same as that of HI clouds. The corresponding minimum, maximum, and mean values of the Galactocentric distances were 9.1 kpc, 18.2 kpc, and 11.3 kpc, respectively. Because the RMS noise in the convolved $^{12}$CO data, velocity resolution, and the maximum heliocentric distance of HI cloud with detected CO emission were $\Delta T_{\rm CO}=0.2$ K, $\Delta v=0.824$ km s$^{-1}$, and $D_{\rm max}=25.1$ kpc, respectively, HI clouds with molecular mass less than $M=\sqrt{125}X_{\rm CO}  T_{\rm min} \Delta v (D_{\rm max} {18\over 3600} {\pi\over 180})^2 {2m_{\rm p}\over M_\odot}=26$ M$_\odot$ could not be counted completely. Similarly to the case of HI, the mass of the faintest detectable H$_2$ gas was estimated to be as small as  5.8 M$_\odot$. It should be noted that H$_2$ clouds without HI gas, referred to as naked H$_2$ clouds, were not considered in our study.

\begin{longtable}{*{11}{c}}
\caption{Identified HI clouds}
\\
\hline
ID & $l$ & $b$ & $V_{\rm LSR}$ & $D$ & $R$ & $r_{\rm min} \times r_{\rm maj}$ & P.A. & $\sigma_v$ & $M_{\rm HI}$ & $M_{{\rm H}_2}$\\
   & [$^\circ$] & [$^\circ$] & [km s$^{-1}$] & [kpc] & [kpc] & [pc $\times$ pc]  & [$^\circ$] & [km s$^{-1}$] & [M$_\odot$] & [M$_\odot$] \\
(1) & (2) & (3) & (4) & (5) & (6) & (7) & (8) & (9) & (10) & (11)\\
\hline
\endfirsthead
\hline
\hline
\endhead
\hline
\endfoot
\hline
\endlastfoot
   1 & $  44.40$ & $  -0.55$ & $-78.1$ &  20.1 &  15.4 &  3.1 $\times$  5.7 & $  -166.8$ &    1.2 &     44.1 $\pm$    1.5 &    182.6 $\pm$    0.2\\
   2 & $  49.00$ & $  -0.45$ & $-76.5$ &  18.1 &  14.2 &  2.9 $\times$  7.9 & $   111.3$ &    2.2 &     41.2 $\pm$    1.3 & ---\\
   3 & $  41.57$ & $  -0.34$ & $-79.8$ &  21.7 &  16.6 &  4.2 $\times$  7.5 & $  -163.9$ &    2.6 &     84.7 $\pm$    2.1 &    288.8 $\pm$    0.3\\
   4 & $  47.60$ & $  -0.29$ & $-78.1$ &  18.8 &  14.7 &  3.3 $\times$  5.6 & $   149.9$ &    2.6 &     85.5 $\pm$    1.8 &    339.6 $\pm$    0.3\\
   5 & $  41.52$ & $  -0.21$ & $-79.8$ &  21.7 &  16.6 &  3.5 $\times$  6.7 & $    66.2$ &    1.4 &     84.0 $\pm$    2.0 &    233.1 $\pm$    0.3\\
   6 & $  49.40$ & $  -0.15$ & $-78.1$ &  18.2 &  14.3 &  2.4 $\times$  5.0 & $   155.8$ &    1.4 &     32.0 $\pm$    1.1 & ---\\
   7 & $  40.81$ & $  -0.03$ & $-76.5$ &  21.3 &  16.1 &  4.0 $\times$  6.1 & $   179.6$ &    2.0 &     44.7 $\pm$    1.6 &    333.4 $\pm$    0.3\\
   8 & $  40.51$ & $   0.17$ & $-75.7$ &  21.3 &  16.0 &  4.0 $\times$  5.0 & $    77.1$ &    1.9 &     68.6 $\pm$    2.0 &    161.1 $\pm$    0.3\\
   9 & $  45.08$ & $   0.24$ & $-78.1$ &  19.8 &  15.2 &  3.8 $\times$  4.9 & $   127.3$ &    1.7 &     79.5 $\pm$    1.4 & ---\\
  10 & $  43.62$ & $   0.54$ & $-79.0$ &  20.6 &  15.8 &  5.8 $\times$ 10.8 & $   136.4$ &    2.0 &    317.9 $\pm$    3.8 &    510.8 $\pm$    0.6\\
  11 & $  40.31$ & $   0.57$ & $-75.7$ &  21.4 &  16.1 &  3.3 $\times$  5.5 & $    63.7$ &    2.0 &     84.8 $\pm$    1.8 & ---\\
  12 & $  44.65$ & $   0.82$ & $-75.7$ &  19.5 &  14.9 &  3.5 $\times$  5.0 & $  -142.6$ &    1.5 &     67.7 $\pm$    1.6 &    110.8 $\pm$    0.2\\
  13 & $  43.15$ & $  -0.22$ & $-72.4$ &  19.5 &  14.7 &  4.8 $\times$  9.6 & $   164.7$ &    2.5 &    575.1 $\pm$    4.0 &    268.0 $\pm$    0.6\\
  14 & $  43.79$ & $  -0.47$ & $-74.8$ &  19.7 &  15.0 &  4.9 $\times$  6.7 & $   150.4$ &    2.5 &    274.9 $\pm$    2.9 &    831.2 $\pm$    0.5\\
  15 & $  49.21$ & $  -0.28$ & $-74.8$ &  17.7 &  13.9 &  4.4 $\times$  7.3 & $   144.5$ &    1.5 &     60.2 $\pm$    1.3 &    281.3 $\pm$    0.2\\
  16 & $  40.78$ & $  -0.06$ & $-76.5$ &  21.3 &  16.1 &  2.5 $\times$  4.2 & $    93.5$ &    1.9 &     37.5 $\pm$    1.5 &     57.8 $\pm$    0.2\\
  17 & $  48.67$ & $   0.77$ & $-74.8$ &  17.9 &  14.0 &  1.8 $\times$  2.7 & $   139.4$ &    1.6 &     67.1 $\pm$    1.1 & ---\\
  18 & $  49.39$ & $  -0.30$ & $-75.7$ &  17.8 &  14.0 &  3.2 $\times$  4.6 & $  -142.9$ &    1.7 &     43.3 $\pm$    1.3 & ---\\
  19 & $  40.29$ & $  -0.02$ & $-74.8$ &  21.2 &  15.9 &  6.3 $\times$  9.8 & $   164.1$ &    1.8 &     67.3 $\pm$    1.9 & ---\\
  20 & $  43.11$ & $   0.05$ & $-76.5$ &  20.3 &  15.4 &  4.9 $\times$  8.9 & $  -137.8$ &    1.3 &    165.3 $\pm$    1.9 &    222.5 $\pm$    0.3\\
\end{longtable}

\section{Properties of identified HI clouds}
\subsection{Mass fraction of HI clouds to the total HI gas}
{While analysis with the dendrogram algorithm can identify clumpy HI clouds, it cannot recover all the diffuse HI component by definition as mentioned in the section 3. Therefore, we first discuss the fraction of HI mass recovered by our analysis to the total mass estimated by integrating the original HI spectra.  

Figure \ref{fig:rgc_MHI} shows the total HI mass, which was estimated by integrating the original HI spectra, and the HI cloud mass, which is estimated by summing up masses of HI clouds identified by the dendrogram algorithm, as a function of the Galactocentric distance. %The total HI mass was estimated by integrating the original HI spactra
The total HI mass was estimated with 
\begin{equation}
M_{\rm HI} = X_{\rm HI} D^2 (\Delta D) m_{\rm p} \int_{b_1}^{b_2}\int_{l_1}^{l_2}\int_{V_1}^{V_2} T_{\rm B}(V_{\rm LSR}) dV_{\rm LSR} dl db,  
\end{equation} 
where $b_1=-1^\circ$, $b_2=+1^\circ$, $l_1=20^\circ$, $l_2=50^\circ$, $D_1=D-\Delta D/2$, $D_2=D+\Delta D/2$, $V_1=V_{\rm LSR} (D_1)$, and $V_2=V_{\rm LSR} (D_2)$. %The HI cloud mass was calculated by summing up the masses of individual HI clouds listed in Table 1. 

The total mass of all HI clouds identified by the dendrogram listed in table 1 amounts to $3.2\times 10^6$ M$_\odot$ while total HI mass in the ranges of $20^\circ\le l \le 50^\circ$ and $R=$9--19 kpc measured $1.6\times 10^7$ M$_\odot$. Therefore, the fraction of the HI emission recovered by the dendrogram analysis was 20 \%. The local fraction varies with the Galactocentric distance $R$, being as high as 23\% around $R=12$ kpc and decreasing below 20\% beyond $R=13$ kpc. This variation seems related to the phase transition between the cold neutral medium (CNM) and the warm neutral medium (WNM) as discussed in the next subsection. 
}

\subsection{Radial variation of HI cloud number density}
Figure \ref{fig:HI_histogram} shows number density of the identified clouds as a function of the Galactocentric distance. This value was determined by counting the clouds within each sector of a Galactocentric radius width of 1 kpc and in a Galactic longitude range of $20^\circ \le l \le 50^\circ$ and then dividing the cloud number by the area in each sector.  

\begin{figure}[t]
\begin{center}
\includegraphics[width=85mm]{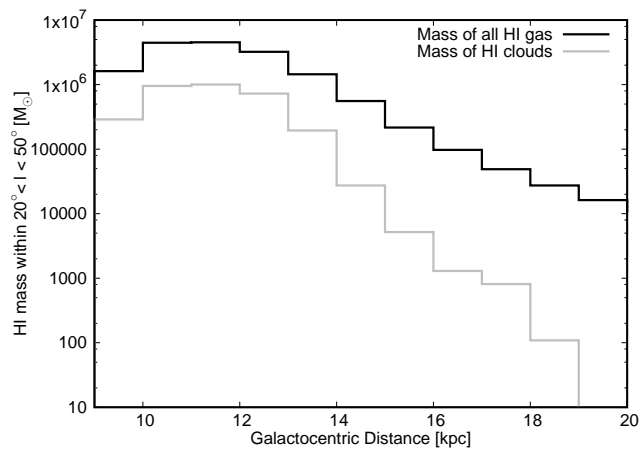}
\end{center}
\caption{Radial distributions of the total HI mass (black line) and HI cloud mass (gray line). The former was calculated by integrating the original HI spectra in a Galactic longitude range of $20^\circ \le l \le 50^\circ$. The latter was calculated by summing up masses of HI clouds listed in table 1. } \label{fig:rgc_MHI}
\end{figure}

\begin{figure}[t]
\begin{center}
\includegraphics[width=85mm]{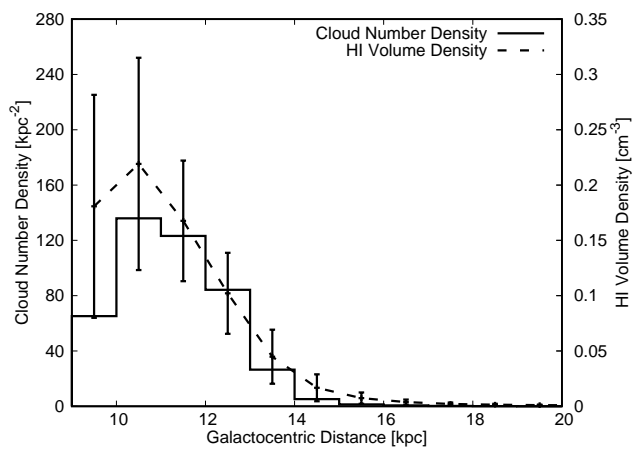}
\end{center}
\caption{Radial distributions of HI cloud number density (solid histogram) and HI volume density (dashed line) in a Galactic longitude range of $20^\circ \le l \le 50^\circ$. The HI cloud number density was calculated on the basis of information given in table 1. The HI volume density was calculated by using the original spectra of $l=20^\circ$--$50^\circ$, $|b|\le 0.5^\circ$. The error bars were derived according to the standard deviation.} \label{fig:HI_histogram}
\end{figure}

The maximum density occurred around a Galactocentric distance of 10 kpc, beyond which the HI cloud number density decreased monotonically with the Galactocentric distance. This peak coincides with locus of the Cygnus arm (Outer arm), which is clearly traced with the HI line, where numerous HI clouds are expected to be identified. 
However, few clouds were identified beyond a Galactocentric distance of 16 kpc in spite of existence of HI gas as shown in figure \ref{fig:rgc_MHI}.

Figure \ref{fig:HI_histogram} additionally shows a comparison of HI cloud number density and HI volume density $n$ at a Galactic latitude of $b=0^\circ$, which was calculated by using the original spectra of $l=20^\circ$--$50^\circ$, $|b|\le 0.5^\circ$ and the aforementioned rotation curve based on the equation 
\begin{equation}
n=X_{\rm HI} T_{\rm HI} \Delta V_{\rm LSR}/\Delta D,  
\end{equation}
where $D$ denotes the heliocentric distance. Because the slopes of both plots were found to be similar, the number of HI clouds is almost proportional to the HI volume density. 

\begin{figure}[ht]
  \begin{center}
\rotatebox{0}{\includegraphics[width=85mm]{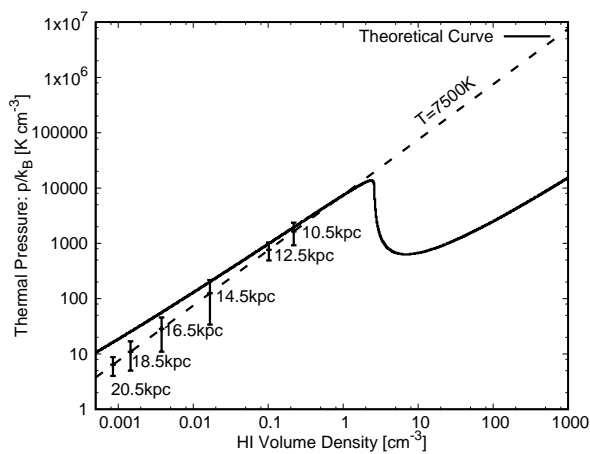}}
  \end{center}
  \caption{Density-pressure diagram. The thick curve is a theoretical model of the thermal equilibrium calculated by using the heating and cooling functions presented by \citet{ino06}. The dashed line shown in the lower HI volume density range of $n<3$ H cm$^{-3}$ corresponds to the WNM phase, and the higher range of $n>10$ H cm$^{-3}$ corresponds to the CNM phase. The mean HI volume densities of each Galactocentric distance derived with original spectra of $b=0$ are plotted as filled squares. } \label{fig:two-phase-model}
\end{figure}

This similarity can be explained by considering the two-phase medium model \citep{wol95} such that clumpy and diffuse HI clouds are CNM and WNM components, respectively. Figure \ref{fig:two-phase-model} shows a thermal equilibrium state of the ISM in the density-pressure diagram. This was calculated on the basis of a model presented by \citet{ino06}, in which the volume density and pressure of the HI gas shown in figure \ref{fig:HI_histogram} are also plotted. The pressure was calculated assuming that the temperature was 7500 K, considering \citet{sof16},who reported a WNM temperature of 7000 K by comparing the observational molecular fraction with a theoretical curve, and \citet{hei03}, who suggested that the temperature of a thermally stable WNM is expected to be 8000 K. This diagram implies that the HI volume density beyond a Galactocentric distance of 16 kpc is lower than 0.01 cm$^{-3}$, and its physical state is quite far from the WNM--CNM transition at about $n=1$--10 cm$^{-3}$. According to \citet{ino06}, the thermal pressure of an ISM with a volume density of 1 cm$^{-3}$ can increase 10--$10^{1.5}$ times due to an external perturbation, and the physical state can move from the WNM phase to the CNM phase. However, the thermal pressure of the HI gas with density of $n<0.01$ cm$^{-3}$ in the outermost region of ($R>16$ kpc) cannot exceed $p/k_{\rm B} = 10^4 $ K cm$^{-3}$ to cross the transition region even if it increases 100 times. Therefore, the HI gas at a Galactocentric distance of $>16$ kpc is considered to remain in the diffuse WNM phase without forming HI clouds or molecular clouds. However, the thermal pressure of the WNM can reach $p/k_{\rm B}\ge 1000$ K cm$^{-3}$ for the inner region with $n\ge 0.1$ cm$^{-3}$, where the phase of ISM can jump into the CNM regime when it is compressed by external perturbation, which leads to the formation of molecules. Such CNM clouds are known as standard clouds \citep{fie69,wol95} and have been likened to raisins in the “raisin pudding” model \citep{cla64,fie69}, although a recent study suggested that these clouds appears in sheets resembling “steamrolled raisins” \citep{kal16}.

\subsection{Volume densities of neutral hydrogen in clouds}
\begin{figure}[ht]
  \begin{center}
\rotatebox{0}{\includegraphics[width=85mm]{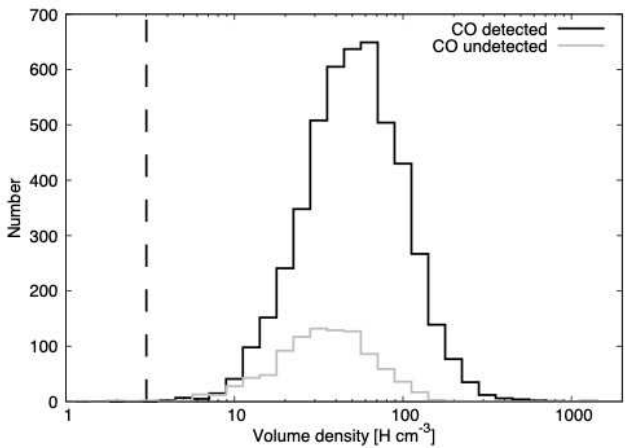}}
  \end{center}
  \caption{ Histograms of volume densities of HI clouds. The vertical line denotes the HI volume density of $n=3$ H cm$^{-3}$, which is the threshold value dividing the phase into the WNM and CNM. HI clouds with and without CO detection are separately plotted with black and gray lines, respectively. } \label{fig:Mtot_density}
\end{figure}

The mean volume densities of neutral hydrogen in the identified HI clouds were calculated by dividing total gas masses $M_{\rm tot}$ by the volume $V={4\pi\over 3}r^3$, where radius of a cloud $r$ was defined as $r=\sqrt{r_{\rm min} r_{\rm maj}}$, as shown in the histogram in figure \ref{fig:Mtot_density}. The mean volume densities of most of the HI clouds (more than 99.7\%) were above the density of $n=3$ H cm$^{-3}$, which is nearly equal to the threshold density between WNM and CNM (figure \ref{fig:two-phase-model}). The fact that most of the clouds have higher gas density than $3$ H cm$^{-3}$ indicates that these clouds are in the CNM phase. However, it should be also mentioned that clouds exist with low mean densities even though the number is small. A possible interpretation is that these are mixtures of CNM- and WNM-phase gases though such clouds are rare.

\subsection{Mass Function}
\begin{figure}[ht]
  \begin{center}
\rotatebox{0}{\includegraphics[width=85mm]{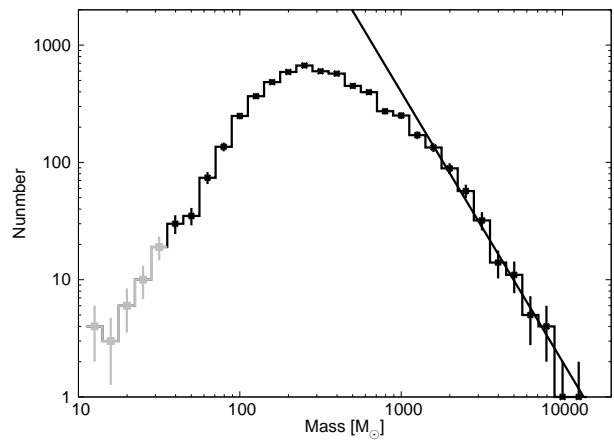}}
  \end{center}
  \caption{The number of the identified HI clouds in each mass interval ($\Delta \log_{10}{M}$) of 0.1. Considering that HI clouds with mass less than $36$ M$_\odot$ were not fully identified, only the mass function for $M>36$ M$_\odot$ meets the completeness and is plotted with a black thick line. The number for the other is incomplete and is plotted with a gray line. The line denotes the power-law function of ${dN\over d(\log_{10}{M})}=10^{9.5} M^{-2.3}=(3.2\times 10^9) M^{-2.3}$.} \label{fig:mass_function_HI}
\end{figure}

Figure \ref{fig:mass_function_HI} shows the mass functions of the HI clouds, which was obtained by counting the number of HI clouds in each mass interval ($\Delta \log_{10}{M}$) of 0.1. Because HI clouds with mass less than 36 M$_\odot$ are incompletely identified, as explained in section 4, the mass function for such HI clouds are plotted with a black line; others are plotted with a gray line. 

First, it should be mentioned that the minimum mass of the HI clouds was 14.0 M$_\odot$, whereas the minimum detectable mass was 8.1 M$_\odot$. This lower limit is close to a mass of WNM within a sphere of acoustic scale $l_a\sim 10$ pc, which is calculated as $l_a=c_s t_c$ using sound speed $c_s$ and cooling timescale $t_c$ \citep{ino06}. This is reasonable because the mass of WNM within a sphere of diameter $l_a$ is estimated as $M=m_{\rm HI} n (4\pi/3)(l_a/2)^3=12$M$_\odot$ assuming an HI density of $n=1$ cm$^{-3}$. However, this lower limit does not match CNM mass within a sphere of CNM at an acoustic scale of $\sim 0.1$ pc or Field length of $\sim 10^{-2}$ pc as well as a WNM mass within a sphere of WNM Field length of $\sim 10^{-1}$ pc. Therefore, the lower limit appears to be consistent with the scenario of HI clouds forming in shocked WNM owing to thermal instability, as suggested by \citet{koy02}.

The HI mass function plotted with the black line in figure \ref{fig:mass_function_HI} can be well fitted with power-law function: 
\begin{equation}
N=N_0 M^{-\alpha}, 
\label{eq:power-law}
\end{equation}
in the mass range of $M \ge 1000$ M$_\odot$ as shown in figure \ref{fig:mass_function_HI}. The factor $N_0$ and power $\alpha$ were estimated to be $N_0=10^{9.5\pm 0.4}$ and $\alpha=2.3\pm 0.1$, respectively. The resultant power of $\alpha=2.3\pm 0.1$ is much larger than that of the Giant Molecular Cloud (GMC), which is known to be typically $\alpha=0.5$ in the Galaxy \citep{sol87}. Because the HI clouds are much smaller in mass than GMCs, it is understandable that they are very different. This steep function seems to imply that most of the atomic mass is within small CNM clouds.

\subsection{H$_2$ mass versus total mass and HI mass}
In order to study the amount of molecular gas formed in each cloud, the H$_2$ mass $M_{{\rm H}_2}$ in each cloud is plotted versus total gas mass $M_{\rm tot}$ in figure \ref{fig:Mtotal-MH2}, which also shows the curves of diffuse and self-gravitating clouds discussed by \citet{elm93}.

In the case of a diffuse cloud with mass of $M$, the molecular mass $M_{{\rm H}_2}$ was calculated with $M_{{\rm H}_2}={{4\pi\over 3}R_{\rm m}^3n\mu}$ using the mean molecular weight $\mu$, the gas number density $n=P/kT$, the interstellar pressure $P$, the Boltzmann constant $k$, the temperature $T$, the radius of the molecular core $R_{\rm m}=R_{\rm e}(1-3S^{-3/2})^{1/3}$, the cloud radius $R_{\rm e}=\sqrt[3]{3M/4\pi n\mu}$, the shielding function $S=(Z\phi_{\rm e}/Z_0 \phi_{{\rm e},0}) (n_e/n_{e,0})^{5/3} (R_e/R_{e,0})^{2/3}$, the metallicity $Z$, and the gas number density $n_{\rm e}$ and radiation intensity $\phi_{\rm e}$ at the edge of a cloud. The subscript 0 indicates the value at the Galactocentric distance of $R=R_0$. The interstellar pressure, metallicity, radiation intensity were taken to be values at $R=R_0$.

In the case of a self-gravitating cloud with mass of $M$, the molecular mass $M_{{\rm H}_2}$ was calculated with $M_{{\rm H}_2}={{4\pi\over 3}R_{\rm m}^3\langle n\rangle \mu}$ using the mean molecular weight $\mu$, the mean gas number density $\langle n\rangle=3n_{\rm e}(R_{\rm e}/R_{\rm m})^2$, the gas number density at the edge of the cloud $n_{\rm e}=217\sqrt{P}/R_{\rm e}$, the radius of the cloud $R_{\rm e}=\sqrt{M/(190\sqrt{P})}$, the interstellar pressure $P$, and the radius of the molecular core $R_{\rm m}$, which was calculated by numerically solving the equation $\left( R_{\rm m}/R_{\rm e}\right)^{-3/2}-\left( R_{\rm m}/R_{\rm e}\right)^{-1/2}=S^{-3/2}$ using the shielding function $S=(Z\phi_{\rm e}/Z_0 \phi_{{\rm e},0}) (n_e/n_{e,0})^{5/3} (R_e/R_{e,0})^{2/3}$, the metallicity $Z$, and radiation intensity $\phi_{\rm e}$ at the edge of the cloud.

Because previous works have suggested that two types of clouds exist, as mentioned above, two sequences are expected to appear in this diagram. Moreover, if molecular gas is rapidly formed in a turbulent ISM on a timescale of 1--2 Myr, as suggested by \citet{glo07}, all clouds would be found along the sequence of self-gravitating clouds because such clouds are H$_2$-dominant.  

However, only a single sequence with large scattering was found, as shown in figure \ref{fig:Mtotal-MH2}. 
The least-squares fit shows that the H$_2$ mass can be expressed with the power-law function $M_{{\rm H}_2}= 10^{-0.75\pm 0.08}M_{\rm tot}^{(1.06\pm 0.03)}$, which implies that the molecular mass is almost linearly proportional to the total gas mass and that typical molecular fraction is $10^{-0.75} \sim 17\%$ of the total gas mass. 

Interestingly, the diagram of $M_{\rm tot}$-$M_{{\rm H}_2}$ shows that the relation between $M_{{\rm H}_2}$ and $M_{\rm tot}$ is scattered near the linear relation with larger dispersion than that of errors. In addition, most of the points are distributed between the two curves of self-gravitating and diffuse clouds. A possible explanation is that each cloud is a mixture of these two clouds, as discussed in the following sections.

\section{Discussion}
\subsection{Possible path of molecular gas evolution}
For more understanding of the $M_{\rm tot}$-$M_{{\rm H}_2}$ diagram, let us compare with a theoretical model shown with three-dimensional magneto-hydrodynamic (MHD) simulation given by \citet{ino12}. They studied molecular cloud formation within a box of $(20 {\rm pc})^3$ with resolution of 0.02 pc. An initial HI medium was set being with fluctuations obeying a power law with the Komorogorov spectral index and was put in a continuous flows with velocity of 20 km s$^{-1}$ along a magnetic field of 5 $\mu$G. A time evolution of molecular clouds were examined by solving the MHD equations, including the effects of chemical reactions, radiative cooling/heating, and thermal conductions. 

The MHD simulations show that diffuse HI gas accumulates onto HI clouds initially formed and that each cloud increases its mass in the flow. According to figure 5 of \citet{ino12}, total mass increases approximately in power-law ($M_{\rm tot}=M_{{\rm tot}_0}(t/t_0)^{\alpha_{\rm tot}}$) from $\sim 4\times 10^3$ M$_\odot$ to $\sim 3\times 10^4$ M$_\odot$ in a time range of 1--10 Myr. Molecular mass also increases approximately in power-law ($M_{{\rm H}_2}=M_{{{\rm H}_2}0}(t/t_0)^{\alpha_{\rm mol}}$) from $\sim 3\times 10^2$ M$_\odot$ to $\sim 2\times 10^4$ M$_\odot$ in the same time range of 1--10 Myr. Therefore, indices $\alpha_{\rm tot}$ and $\alpha_{{\rm H}_2}$ are $\log_{10}{(3\times10^4/4\times10^3)}=0.9$ and $\log_{10}{(2\times10^4/3\times10^2)}=1.8$, respectively. Therefore, it is found that the mass of molecular clouds evolves as $M_{{\rm H}_2}\propto M_{\rm tot}^{\alpha_{{\rm H}_2}/\alpha_{\rm tot}}\propto M_{\rm tot}^2$ by eliminating $t$ from these expressions, though this relation is not explicitly mentioned in \citet{ino12}.

\begin{figure}[h]
  \begin{center}
  \rotatebox{0}{\includegraphics[width=85mm]{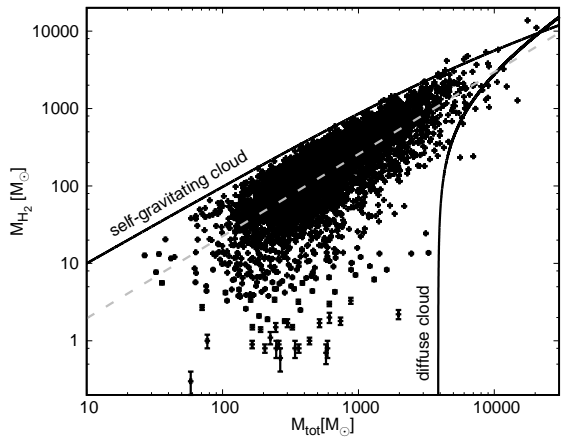}}
  \end{center}
\caption{H$_2$ mass versus total mass including HI and H$_2$ gas masses. The power-law function $M_{{\rm H}_2} = 10^{-0.75} M_{\rm tot}^{1.06}=0.17 M_{\rm tot}^{1.06}$, obtained by least-squares fitting, is plotted with a gray dashed line. The curves of the theoretical model based on \citet{elm93} are also plotted and labeled as self-gravitating cloud and diffuse cloud. } \label{fig:Mtotal-MH2}
\end{figure}

\begin{figure}[h]
  \begin{center}
  \rotatebox{0}{\includegraphics[width=85mm]{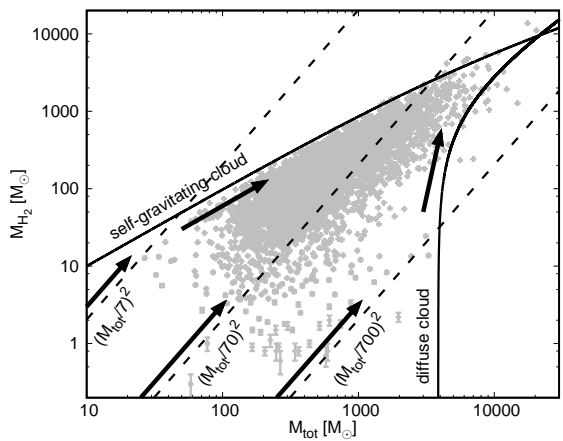}}
  \end{center}
  \caption{Possible paths of molecular cloud evolution superposed on the image shown in figure \ref{fig:Mtotal-MH2}. The dashed lines labeled $(M_{\rm tot}/7)^2$, $(M_{\rm tot}/70)^2$, and $(M_{\rm tot}/700)^2$ denote paths predicted by \citet{ino12}. } \label{fig:possible-path}
\end{figure}

Considering the observations mentioned above, figure \ref{fig:possible-path} suggests possible paths along which clouds evolve and experience increases in molecular mass. In this figure, the same data as those shown in figure \ref{fig:Mtotal-MH2} are plotted with gray dots; the cases of self-gravitating and diffuse clouds are shown by solid curves; and the predicted evolution paths of $M_{{\rm H}_2} = (M_{\rm total}/M_0)^2$ as described above, where $M_0$ denotes the total gas mass of a cloud containing molecular mass of $M_{{\rm H}_2}=1$ M$_\odot$, are also overlain by dashed lines. Three cases of mass are presented as examples: $M_0=7$ M$_\odot$, 70 M$_\odot$, and 700 M$_\odot$. 

If the molecular fraction of clouds is less than unity, the mass of molecular gas increases along the lines of $M_{{\rm H}_2} = (M_{\rm HI}/M_0)^2$, as indicated in figure \ref{fig:possible-path} by the three arrows aligned with dashed lines. If a cloud reaches the self-gravitating cloud phase by increasing the molecular mass, the cloud would evolve along the line of the self-gravitating phase with no further increase in the molecular mass. This path is indicated in the figure by the arrow aligned with the self-gravitating cloud line. Similarly, if a cloud reaches the curve of the diffuse cloud phase, the molecular fraction will increase rapidly, as indicated by the arrow aligned with the curve of diffuse cloud phase.

\subsection{HI and CO distributions of the four most massive HI clouds}
\begin{figure*}[h]
  \begin{center}
  \includegraphics[width=170mm]{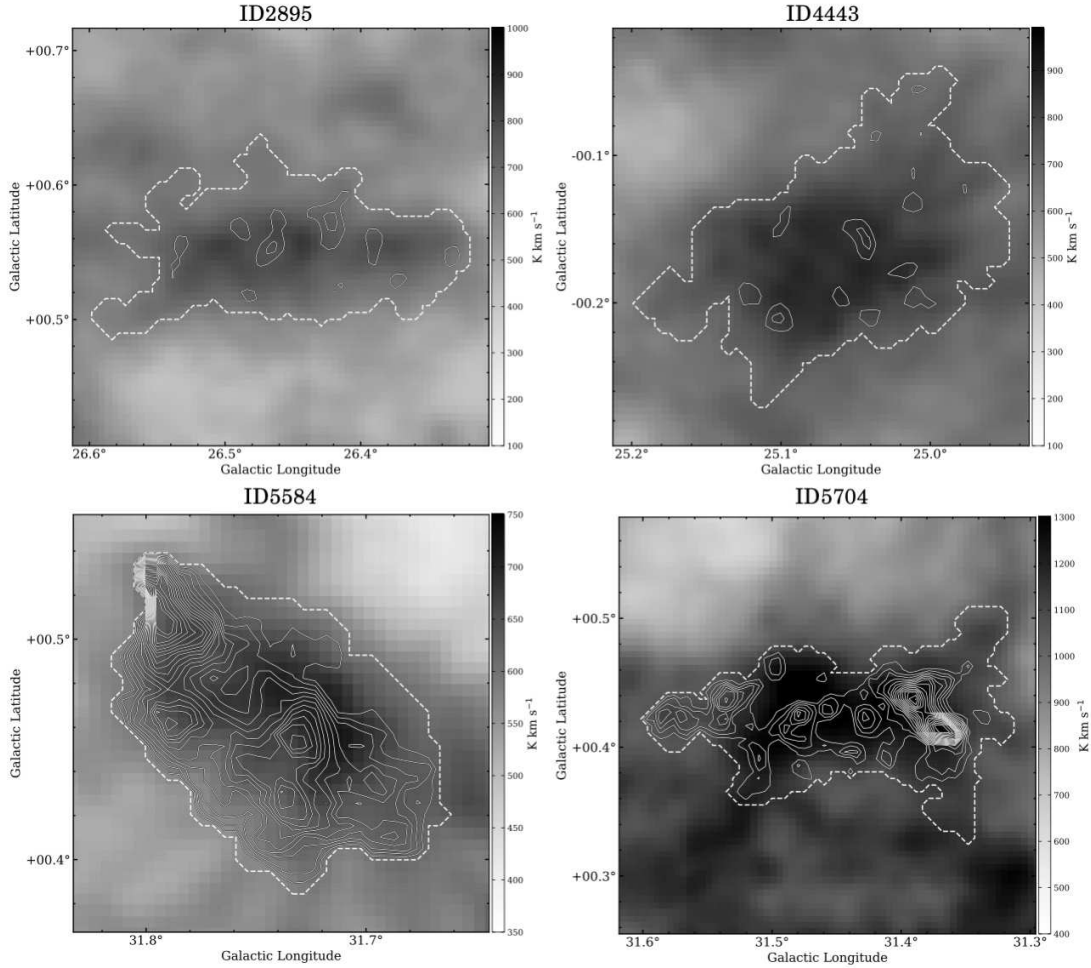}
  \end{center}
  \caption{Distributions of HI gas (gray scale) and H$_2$ gas (contours) in the four most massive HI clouds.  The contour levels are 1, 2, 3,...17 K km s$^{-1}$. The white thick dashed lines denote the boundaries of HI clouds identified by using the dendrogram. The identification numbers correspond to those of the clouds listed in table 1.} \label{fig:HIintegmaps}
\end{figure*}

Figure \ref{fig:HIintegmaps} shows the HI and H$_2$ distributions in the four most massive HI clouds, which are presented by gray images and contours, respectively. The thick dashed lines indicate the edges of the HI clouds identified by using the dendrogram. The same physical parameters shown in table 1 are listed in table 2 for these four HI clouds. As shown in figure \ref{fig:HIintegmaps}, the HI gas was distributed asymmetrically in all cases. %, and the molecular gas could be detected at the edges as well as the centers of the HI clouds.  

Previous research indicates that ISM clouds consist of molecular cores at the centers that are surrounded by spherical HI envelopes, resembling jam-doughnuts \citep{elm93,kru08,mck10,ste14}. However, they actually consist of small molecular clouds distributed over large HI clouds, resembling chocolate-chip scones.

Although we do not consider the theoretical model given by previous research to be vastly different from actual conditions because the model has been able to roughly explain the observations thus far, we suggest that the model requires modification.  
Considering that the relation between molecular gas and total gas mass is restricted within the range of theoretically predicted self-gravitating and diffuse clouds, as discussed in the previous section, a large molecular cloud likely consists of an assemblage of smaller self-gravitating clouds. 

As shown in figure \ref{fig:HIintegmaps}, clouds 2895 and 4443 exhibit the most typical characteristics of the aforementioned chocolate-chip scone model.  %In the case of cloud 7846, the molecular gas distribution is one-sided to the west. This indicates that the cloud previously collided with another cloud or experienced ram pressure owing to gas flow to form molecular gas in a triggered shock front. 
Clouds 5584 and 5704 contain more molecular gas, which can be seen at the edges as well as the centres of the HI clouds. This seems to indicate that the cloud previously collided with another cloud or experienced ram pressure owing to gas flow to form molecular gas in a triggered shock front.

\begin{figure}[h]
  \begin{center}
  \includegraphics[width=80mm]{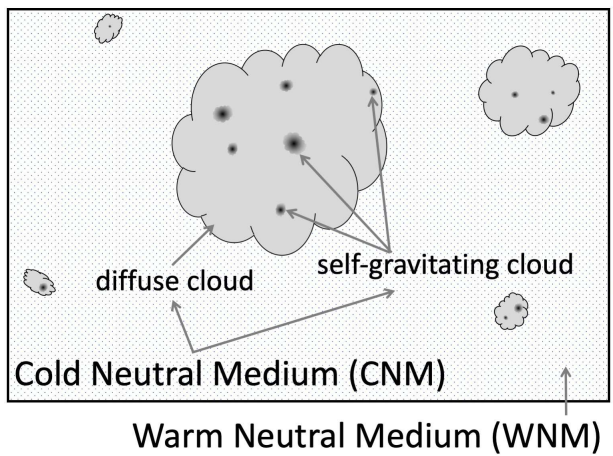}
  \end{center}
  \caption{Schematic diagram of the chocolate-chip scone model. The black and gray clouds are self-gravitating and diffuse clouds, respectively, both of which are in the CNM phase. Most of the clouds are composed of a mixture of these two cloud types. The CNM clouds, resembling chocolate-chip scones, are distributed within a “soup” of WNM. } \label{fig:schematicview}
\end{figure}

These images effectively illustrate the chocolate-chip scone model, in which the CNM cloud consists of H$_2$-dominant small self-gravitating clouds (chocolate-chips) and cold HI-dominant diffuse clouds (plain scone). 
Both self-gravitating and diffuse clouds are essentially in the CNM phase because the volume density is in the range of the CNM phase, as shown in subsection 5.3. Such CNM clouds are distributed inside a “soup” of HI-dominant WNM. A schematic diagram of the chocolate-chip scone model described above is given in figure \ref{fig:schematicview}.

\begin{longtable}{*{11}{c}}
\caption{Four most massive HI clouds}
\\
\hline
ID & $l$ & $b$ & $V_{\rm LSR}$ & $D$ & $R$ & $r_{\rm min} \times r_{\rm maj}$ & P.A. & $\sigma_v$ & $M_{\rm HI}$ & $M_{{\rm H}_2}$\\
   & [$^\circ$] & [$^\circ$] & [km s$^{-1}$] & [kpc] & [kpc] & [pc $\times$ pc]  & [$^\circ$] & [km s$^{-1}$] & [M$_\odot$] & [M$_\odot$] \\
(1) & (2) & (3) & (4) & (5) & (6) & (7) & (8) & (9) & (10) & (11)\\
\hline
\endfirsthead
\hline
\hline
\endhead
\hline
\endfoot
\hline
\endlastfoot
2895 & $  26.44$ & $   0.55$ & $-35.3$ &  18.8 &  12.1 &  7.2 $\times$ 18.8 & $  -177.7$ &    2.4 &  10413.0 $\pm$    6.5 &   1922.7 $\pm$    1.0\\
4443 & $  24.99$ & $  -0.17$ & $-29.5$ &  18.2 &  11.4 & 10.2 $\times$ 16.7 & $  -148.0$ &    2.0 &  13549.1 $\pm$    6.4 &   1270.1 $\pm$    1.0\\
5584 & $  31.41$ & $   0.44$ & $-20.4$ &  15.5 &   9.6 &  6.8 $\times$ 16.2 & $  -175.6$ &    3.3 &   9370.3 $\pm$    4.7 &  11061.5 $\pm$    0.8\\
5704 & $  31.72$ & $   0.46$ & $-20.4$ &  15.5 &   9.6 &  5.8 $\times$ 10.0 & $   139.7$ &    1.7 &   3819.5 $\pm$    3.1 &  13741.6 $\pm$    0.5\\
\end{longtable}

\subsection{Possible Systematic Errors in the Analysis}
\begin{figure*}[h]
  \begin{center}
  \rotatebox{0}{\includegraphics[width=80mm]{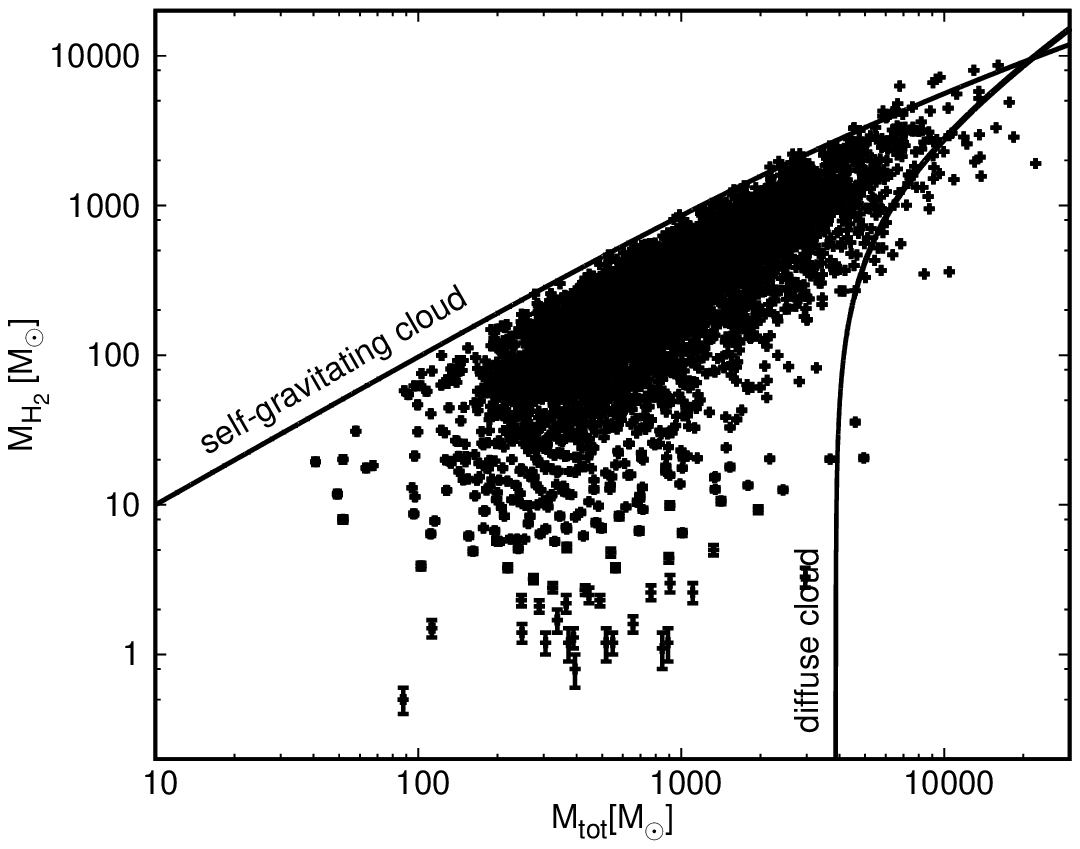}}
  \rotatebox{0}{\includegraphics[width=80mm]{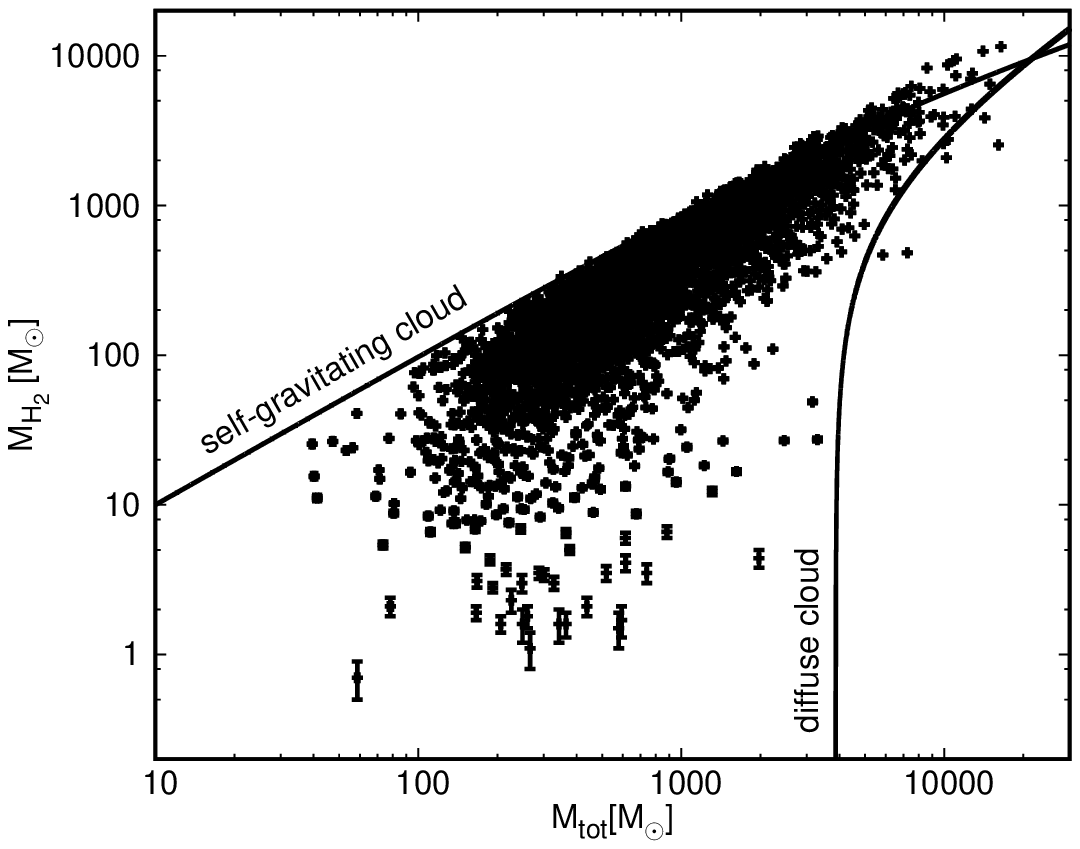}}
  \end{center}
\vspace*{1.cm} 
\caption{$M_{\rm tot}$-$M_{{\rm H}_2}$ diagrams in adopting another kinematic distance (left) and another CO-to-H$_2$ conversion factor (right). } \label{fig:Mtotal-MH2-SysErr}
\end{figure*}
In addition to the errors due to random noise in the observational data considered above, we discuss how possible systematic errors, such as kinematic distance and CO-to-H$_2$ conversion factors, affect the main result shown in figure \ref{fig:Mtotal-MH2}. As shown below, these systematic errors do not change the appearance of $M_{\rm tot}$-$M_{{\rm H}_2}$ diagram and does not affect our result. 

\subsubsection{Kinematic distance}
Heliocentric distance $D$ of each cloud was calculated based on the kinematic distance, which depends on a choice of the rotation curve and the Galactic constants $R_0$ and $V_0$. These would systematically affect the estimation of mass by factor of $D^2$. In order to test how much it affects the $M_{\rm tot}$-$M_{{\rm H}_2}$ diagram, let us check the case of adopting a flat rotation curve and another set of Galactic constants $R_0=10$ kpc and $V_0=250$ km s$^{-1}$. Left panel of figure \ref{fig:Mtotal-MH2-SysErr} shows $M_{\rm tot}$-$M_{{\rm H}_2}$ diagram in adopting this kinematic distance. Whereas the kinematic distance is estimated to be larger, figure \ref{fig:Mtotal-MH2-SysErr} is almost the same appearance as figure \ref{fig:Mtotal-MH2} because both HI and H$_2$ masses are estimated largely and plot moves upward from left to right in the diagram but does not change the appearance.

\subsubsection{CO-to-H$_2$ conversion factor}
The H$_2$ mass depends on a choice of CO-to-H$_2$ conversion factor. While it is suggested that the conversion factor varies with the Galactocentric distance \citep{ari96}, we adopted constant conversion factor in the above discussions. Therefore, similarly to the case of kinematic distance, let us test how much the choice of conversion factor affects the $M_{\rm tot}$-$M_{{\rm H}_2}$ diagram. Right panel of figure \ref{fig:Mtotal-MH2-SysErr} shows $M_{\rm tot}$-$M_{{\rm H}_2}$ diagram in adopting two times larger conversion factor. Similarly to the previous case, the appearance of figure \ref{fig:Mtotal-MH2-SysErr} is almost the same as figure \ref{fig:Mtotal-MH2} because the conversion factor make both H$_2$ and total masses larger and plot moves upward from left to right in the diagram but does not change the appearance.

\section{Conclusion}
We applied dendrogram analysis to the VGPS HI data cube to create a catalog of HI clouds of the outer Galaxy, which includes masses of H$_2$ gas within each HI cloud that were estimated by using FUGIN CO survey data. The ranges of the detected HI and H$_2$ masses are 14.0--$1.35\times 10^4$ M$_\odot$ and 0.300--$1.37\times 10^4$ M$_\odot$, respectively.  
The radial distribution of the HI cloud number density correlated strongly with the HI volume density, and most of the identified HI clouds were found to be in the CNM phase. 
The high-mass end of the mass function of the HI clouds fit well with the power-law function, with index of 2.3. 
Previous theoretical studies based on the jam-doughnut model indicate that clouds can be divided into two sequences of self-gravitating and diffuse. However, the M$_{\rm tot}$-M$_{{\rm H}_2}$ diagram shows only a single sequence with large scattering rather than two sequences. Considering that the M$_{\rm tot}$-M$_{{\rm H}_2}$ relation of each cloud is found between these two sequences, we suggest that most of the clouds can be modeled by using the chocolate-chip scone model, which contains a mixture of self-gravitating clouds resembling chocolate chips and diffuse clouds resembling a scone. This description is consistent with the images shown in the CO and HI maps of the four most massive clouds. 

Considering that H$_2$ gas increases with $M_{{\rm H}_2} \propto M_{\rm tot}^2$, as implied in a recent work, we suggest that the molecular clouds first evolved along the path of $M_{{\rm H}_2} \propto M_{\rm tot}^2$ by collecting diffuse gas and then moved along the curves of self-gravitating and diffuse clouds.

\noindent{\bf Acknowledgement} We thank the FUGIN project members for conducting observations. We would also like to thank anonymous referee for carefully reading the manuscript.

\end{document}